\def\be{\begin{equation}}
\def\ee{\end{equation}}
\def\ba{\begin{array}}
\def\ea{\end{array}}

\def\Rb{{I\!\! R}}

\def\Cb{\ \hbox{\vrule width 0.6pt height 6pt depth 0pt
		      \hskip -3.2 pt} C}
\documentstyle[12pt]{article}
\begin{document}
\parskip=4pt
\parindent=18pt
\baselineskip=22pt
\setcounter{page}{1}
\centerline{\Large\bf A Note on Invariants and Entanglements}
\vspace{6ex}
\begin{center}
{\large  Sergio Albeverio}\footnote {SFB 256; BiBoS;
CERFIM (Locarno); Acc.Arch., USI (Mendrisio)}~~~ and ~~~
{\large  Shao-Ming Fei}\footnote{Institute of Applied Mathematics,
Chinese Academy of Science, Beijing.}
\end{center}
\begin{center}
Institut f\"ur Angewandte Mathematik, Universit\"at Bonn,
D-53115 Bonn
\end{center}
\vskip 1 true cm
\parindent=18pt
\parskip=6pt
\begin{center}
\begin{minipage}{5in}
\vspace{3ex}
\centerline{\large Abstract}
\vspace{4ex}
The quantum entanglements are studied in terms of the
invariants under local unitary transformations.
A generalized formula of concurrence for $N$-dimensional
quantum systems is presented. This generalized concurrence
has potential applications in studying separability
and calculating entanglement of formation for high dimensional mixed
quantum states.

\bigskip
\medskip
\bigskip
\medskip

\vfill

\end{minipage}
\end{center}

\newpage

Quantum entanglement is tightly related to the foundations of 
quantum mechanics, particularly to quantum nonseparability and the
violation of Bell's inequalities \cite{Bell}.  
It has also been playing important roles in communication,
information processing and quantum computing \cite{DiVincenzo}, such as
in the investigation of quantum teleportation \cite{teleport,teleport1},
dense coding \cite{dense},
decoherence in quantum computers and
the evaluation of quantum cryptographic schemes \cite{crypto}.
To quantify entanglement, a well justified and mathematically 
tractable measure is needed. A number of entanglement measures
such as the entanglement of formation and distillation
\cite{Bennett96a,BBPS,Vedral},
negativity \cite{Peres96a,Zyczkowski98a}, von Neumann entropy
and relative entropy \cite{Vedral,sw} have been proposed
for bipartite states [6,8,12-15] and some of their relations
have been discussed \cite{sr}, though most proposed measures of
entanglement involve extremizations which
are difficult to handle analytically. 

The entanglement of formation \cite{Bennett96a} is intended to
quantify the amount of quantum communication required to create a
given state. For the entanglement of a pair of qubits, it has been shown
that the entanglement of formation can be expressed as a 
monotonically increasing function of the ``concurrence'' $C$. This
function ranges from 0 to 1 as $C$ goes from 0 to 1, 
so that one can take the
concurrence as a measure of entanglement in its own right
\cite{HillWootters}. From the expression of $C$, 
which is much simpler than
the definition of entanglement of formation, the entanglement of formation
for mixed states of a pair of qubits is calculated \cite{HillWootters}. 
Nevertheless so far no explicit analytic formulae 
for entanglement of formation have been found for systems larger than a
pair of qubits (the case being special in many ways \cite{ww}), 
although entanglement of formation is 
defined for arbitrary dimension. 

In fact, as the degree of entanglement will neither
increase nor decrease under local unitary transformations
on a subquantum system, the
measure of entanglement must be an invariant of local unitary
transformations. In this note we describe entanglements from
the view of this kind of invariants. A generalized explicit
formula of concurrence for high dimensional bipartite systems is derived
from the relations among these invariants.

Consider the case of quantum systems with an
$N$-dimensional complex Hilbert space ${\cal H}$.
Let $e_i$, $i=1,...,N$, be an orthonormal the basis of the Hilbert space.
A general pure state of two $N$-dimensional quantum systems is of the form,
\begin{equation}
\vert\Psi>=\sum_{i,j=1}^N a_{ij}e_i\otimes e_j,~~~~~~a_{ij}\in\Cb
\end{equation}
with normalization 
\begin{equation}\label{norm}
\sum_{i,j=1}^N a_{ij}a_{ij}^\ast=1\,.
\end{equation}
The entanglement of formation is given by
$$
E(\vert\Psi>)=-Tr(\rho_0\log_2\rho_0),
$$
where $\rho_0$ is the partial trace of $\vert\Psi><\Psi\vert$ over one of the
subsystems. For $N=2$, the state $\vert\Psi>$ is
factorizable into single-qubit (unentangled) states
if and only if $a_{11}a_{22} = a_{21}a_{12}$. It is shown that
\be\label{c2}
C=2|a_{11}a_{22}-a_{12}a_{21}\vert,
\ee
which ranges from 0 to 1, is a plausible measure of the degree of 
entanglement. This is taken to be the
definition of \underline{concurrence} 
for a pure state of two qubits \cite{HillWootters}.

Let $U$ denote the unitary transformations 
on the Hilbert space ${\cal H}$, such that
\begin{equation}\label{unit}
U e_i \mapsto \sum_{j=1}^N b_{ij}e_j,~~~~~~b_{ij}\in\Cb
\end{equation}
and
$$
\sum_{j=1}^N b_{ij}b_{kj}^\ast=\delta_{ik}.
$$
We call a quantity an invariant associated with the state
$\vert\Psi>$ if it is invariant under the local unitary transformations of
$U\otimes U$. Let $A$ denote the matrix given by $(A)_{ij}=a_{ij}$.
Hence $\rho_0=AA^\dagger$.
By generalizing the results of analysis on invariants
for qubits \cite{Linden}, we can show that the following quantities
are invariants under local unitary transformations:
\begin{equation}\label{I}
I_\alpha=Tr(AA^\dag)^{\alpha+1},~~~~~~~~~~~\alpha=0,1,...,N-1.
\end{equation}
Among these invariants,
\begin{equation}\label{I0}
I_0=Tr(AA^\dag)=\sum_{i,j=1}^N a_{ij}a_{ij}^\ast
\end{equation}
and
\begin{equation}\label{I1}
I_1=Tr[(AA^\dag)^2]=
\sum_{i,j,k,m=1}^N a_{ik}a_{im}^\ast a_{jm}a_{jk}^\ast,
\end{equation}
are of particular importance.
$I_0$ is in fact the normalization condition (\ref{norm}).
$I_1$ is a biquadratic form of the coefficients
$a_{ij}$.

Polynomials in $I_0$ and $I_1$ are obviously also invariants. Among
them the quantity $\frac{N}{N-1}(I_0^2-I_1)$ is of special significance:
$$
\ba{rcl}
\displaystyle\frac{N}{N-1}(I_0^2-I_1)&=&\displaystyle\frac{N}{N-1}
\sum_{i,j,k,m=1}^N (a_{ik}a_{ik}^\ast a_{jm}a_{jm}^\ast
-a_{ik}a_{im}^\ast a_{jm}a_{jk}^\ast)\\[4mm]
&=&\displaystyle\frac{N}{2(N-1)}\sum_{i,j,k,m=1}^N 
(a_{ik}a_{jm}-a_{im}a_{jk})
(a_{ik}^\ast a_{jm}^\ast-a_{im}^\ast a_{jk}^\ast)\\[4mm]
&=&\displaystyle\frac{N}{2(N-1)}
\sum_{i,j,k,m=1}^N \vert a_{ik}a_{jm}-a_{im}a_{jk}\vert^2.
\ea
$$
For the case of $N=2$, we see that the 
square root of $\frac{N}{N-1}(I_0^2-I_1)$ is just the concurrence
(\ref{c2}). For general $N$, we see that 
$\frac{N}{N-1}(I_0^2-I_1)$ is positive
definite and takes values from $0$ to $1$. It takes the value
zero when the state $\vert\Psi>$ is factorizable, $a_{ij}=a_ib_j$ for some
$a_i$, $b_j\in\Cb$, $i,j=1,...,N$, 
and one when $\vert\Psi>$ is maximally entangled,
e.g., $a_{ii}=1/\sqrt{N}$, $a_{ij}=0$ for $i \neq j$.
The term $\vert a_{ik}a_{jm}-a_{im}a_{jk}\vert$
stands for the contribution of the terms $a_{ik}e_i\otimes e_k+
a_{jm}e_j\otimes e_m+a_{im}e_i\otimes e_m+a_{jk}e_j
\otimes e_k$ in $\vert\Psi>$
to the ``degree of entanglement". When $a_{ik}a_{jm}-a_{im}a_{jk}=0$, these
terms can be written in a factorized form 
$(a_1 e_i+ a_2 e_j)\otimes(b_1 e_k + b_2 e_m)$ 
for some $a_1,b_1,a_2,b_2\in \Cb$ and
give zero contribution to the entanglement.
When $a_{ik}a_{jm}=1$ (resp. $a_{im}a_{jk}=1$) and
$a_{im}a_{jk}=0$ (resp. $a_{ik}a_{jm}=0$), they give the maximal
contribution to the entanglement. If $\vert\Psi>$ is unentangled,
i.e., $\vert\Psi>=
(\sum_{i=1}^N a_{i}e_i)\otimes (\sum_{j=1}^N b_{j}e_j)$,
$a_{ij}=a_i b_j$, then $\vert a_{ik}a_{jm}-a_{im}a_{jk}\vert=0$,
$\forall i,j,k,m =1,...,N$, and we have $I_0^2-I_1=0$. All entangled states
have at least one of the terms $\vert a_{ik}a_{jm}-a_{im}a_{jk}\vert\neq
0$, hence they have non zero $I_0^2-I_1$.

Therefore $\frac{N}{N-1}(I_0^2-I_1)$ could be a suitable candidate for the
measure of entanglement of two $N$ dimensional quantum systems
$\vert\Psi>$ in some sense. 
In accord with the definition of concurrence for two qubits
\cite{HillWootters}, we take
the square root of it to be the generalized concurrence:
\be\label{cn}
C_N=\sqrt{\frac{N}{N-1}(I_0^2-I_1)}=\sqrt{\displaystyle
\frac{N}{2(N-1)}\sum_{i,j,k,m=1}^N \vert
a_{ik}a_{jm}-a_{im}a_{jk}\vert^2}\,.
\ee
It is seen that $C_2=C$. The factor $\frac{N}{N-1}$ in (\ref{cn}) is
just so chosen such that the maximal value of $C_N$ is scaled to be one.

To understand the relations between the generalized concurrence $C_N$
and the 
entanglement of formation $E(\vert\Psi>)$, 
we now study the properties of the 
invariants. In terms of the Schmidt decomposition, a given
$\vert\Psi>$ can always be written in the form, in 
some orthonormal basis $\{e_i\}$, $i=1,...,N$,
$$
\vert\Psi>=\sum_{i=1}^N \sqrt{\Lambda_i}e_i\otimes e_i,
$$
where $\sum_{i=1}^N \Lambda_i=1$, $\Lambda_i\geq 0$. The matrix
$\rho_0$ is then of the form $\rho_0=AA^\dag=diag(\Lambda_1,...\Lambda_N)$.
The entanglement of formation of $\vert\Psi>$ is given by 
\be\label{it}
E(\vert\Psi>)=-\sum_{i=1}^N \Lambda_i\log_2\Lambda_i.
\ee
The invariants are then of the form
$$
I_\alpha=\sum_{i=1}^N \Lambda_i^{\alpha +1},~~~~~~~\alpha=0,...,N-1.
$$

First we note that
$$
I_1=\sum_{i=1}^N \Lambda_i^2=I_0^2-
\sum_{i \neq j}^N \Lambda_i\Lambda_j.
$$
Therefore $C_N=0$ implies that $\sum_{i \neq j}^N \Lambda_i \Lambda_j=0$.
As $\Lambda_i\geq 0$ and $\sum_{i=1}^N \Lambda_i=1$, we have that only 
one $\Lambda$,
say $\Lambda_1$, equals to $1$ and the rest be zero. In this case
$I_\alpha=I_0^{\alpha+1}$, $\alpha=1,...,N-1$, and $E(\Psi)=0$.
That is $I_0^2-I_1=0$ implies $E(\Psi)=0$.

If $C_N=1$, we have $\sum_{i \neq j}^N \Lambda_i \Lambda_j=\frac{N-1}{N}$,
which is equivalent to the condition
$\sum_{i=1}^N \Lambda_i^2=1/N$, according to the normalization
$\sum_{i=1}^N \Lambda_i=1$.
Equation $\sum_{i=1}^N \Lambda_i^2=1/N$ describes a $N-1$-dimensional
sphere in $\Rb^N$ with radius $1/\sqrt{N}$, whereas $\sum_{i=1}^N 
\Lambda_i=1$
is a hyperplane in $\Rb^N$. These geometrical objects 
have only one contact point at 
$\Lambda_i=1/N$ for $\Lambda_i\geq 0$, $i=1,...,N$. Therefore $C_N=1$
implies that $E(\Psi)=\sum_{i=1}^N \Lambda_i\log_2\Lambda_i=1$.

Therefore the invariant $I_0^2-I_1$ and hence the generalized 
concurrence $C_N$ characterizes the properties of entanglement in
some way. However
we remark that the above properties of $C_N$ do not mean that
$C_N$ is in general a suitable measure for general $N$-dimensional
bipartite quantum pure states. It can however be shown that when the
matrix $AA^\dag$ has only two different nonzero eigenvalues, the
entanglement of formation is a monotonically
increasing function of $C_N$, thus  $C_N$ can indeed be used as 
a measure of entanglement in this case.

In fact one can show that the 
eigenvalue equation for the $\Lambda_i$, $i=1,...,N$, has the form:
$$
\sum_{k=1}^N(-\Lambda)^k c_k=0
$$
with $c_k$ polynormials in the invariants $I_0,...,I_{N-1}$. E.g.,
\begin{equation}\label{cc}
\begin{array}{l}
c_N=1,~~~~~ c_{N-1}=I_0,~~~~~
c_{N-2}=\displaystyle\frac{1}{2}(I_0^2-I_1),\\[4mm]
c_{N-3}=\displaystyle\frac{1}{6}(I_0^3+2 I_2-3 I_0 I_1),\\[4mm]
c_{N-4}=\displaystyle\frac{1}{24}(I_0^4-6 I_0^2 I_1+8I_0I_2+3I_1^2-6I_3).
\end{array}
\end{equation}
As $I_0$ is normalized to be one, the coefficient $c_{N-2}=(I_0^2-I_1)/2$
is the first non trivial one in the eigenvalue equation of $\Lambda$. 
To see the role of $I_0^2-I_1$ in the equation, let us take $N=3$. We have
then
$$
\begin{array}{l}
\Lambda_1=\displaystyle
\frac{1}{3}+\frac{2}{3}\cos\frac{\phi}{3}\sqrt{1-C_3^2}\\[4mm]
\Lambda_2=\displaystyle\frac{1}{3}-\frac{1}{3}(\cos\frac{\phi}{3}
+\sqrt{3}\sin\frac{\phi}{3})\sqrt{1-C_3^2}\\[4mm]
\Lambda_3=\displaystyle\frac{1}{3}-\frac{1}{3}(\cos\frac{\phi}{3}
-\sqrt{3}\sin\frac{\phi}{3})\sqrt{1-C_3^2}\\[4mm]
\end{array}
$$
where $C_3^2=(C_3)^2=\frac{3}{2}(I_0^2-I_1)$, $\phi=\arctan
\frac{\sqrt{B_2}}{B_1}$, $B_1=2-9 c_1 + 27 c_0$, $B_2=
\vert 4 (3c_1-1)^3 + B_1\vert$, $c_1,c_2$ as in (\ref{cc}).
For $C_3=0$ (resp. $C_3=1$) we have $\Lambda_1=1$, 
$\Lambda_2=\Lambda_3=0$
(resp. $\Lambda_1=\Lambda_2=\Lambda_3=1$).
Nevertheless, the entanglement of formation (\ref{it}) is not a
monotonically increasing function of $C_3$. This is different
to the case $N=2$, where according to the condition
$\sum_{i=1}^N \Lambda_i=1$, there is only one independent
eigenvalue of $\rho_0$.

The generalized concurrence (\ref{cn}) is useful in finding the
necessary and sufficient conditions for the separability of 
mixed states and in caculating the entanglement of formation
for some classes of mixed desity matrices.
Due to recent works by Peres \cite{Pe96} and 
Horodecki et al \cite{Ho96} there exist a simple
criterion allowing one to judge, whether a given density matrix $\rho$, 
representing a $2 \times 2 $ or $2 \times 3$ composite system, is
separable. Nevertheless,
the general problem of finding sufficient and necessary conditions for
separability in higher dimensions remains open 
(see e.g. \cite{LBCKKSST,H300} and references therein).
A general condition for separability of a quantum state
could  in principle be obtained from
the measure of entanglement. However most proposed measures of
entanglement involve extremizations which
are difficult to handle analytically.
For instance, the ``entanglement of formation'' \cite{Bennett96a},
is defined for arbitrary dimension, but  so far no
explicit analytic formulae for entanglement of formation
have been found for systems larger than a
pair of qubits (spin-$1\over 2$ particles).
As applications of the generalized concurrence, we have presented
the necessary and sufficient conditions for the separability of 
high dimensional rank two mixed states \cite{qsep}.
Let $\rho$ be a rank two state in ${\cal H}\otimes {\cal H}$, with
$|E_1\rangle$, $|E_2\rangle$ being its two orthonormal eigenvectors
corresponding to the two nonzero eigenvalues:
\be\label{rho}
\rho=p|E_1\rangle\langle E_1| +q|E_2\rangle\langle E_2|,
\ee
where $q=1-p\in (0,1)$.
Generally $|E_k\rangle=
\displaystyle\sum_{i,j=1}^N a_{ij}^k e_i\otimes e_j$,
$a_{ij}^k\in\Cb$,
with normalization 
$\displaystyle\sum_{i,j=1}^N a_{ij}^k (a_{ij}^{k})^\ast=1$, $k=1,2$.
With the notations:
$$
\ba{l}
\alpha_{ij}^{kl}=a^2_{ij}a^2_{kl}-a^2_{il}a^2_{kj},~~~
\gamma_{ij}^{kl}=a^1_{ij}a^1_{kl}-a^1_{il}a^1_{kj}\\[3mm]
\beta_{ij}^{kl}=a^1_{ij}a^2_{kl}+a^2_{ij}a^1_{kl}-a^2_{il}a^1_{kj}
-a^1_{il}a^2_{kj},
\ea
$$
it is shown that
$\rho$ is separable if and only if there is $\theta\in\Rb$ such that  
\be\label{t21}
\gamma_{ij}^{kl}=e^{i\theta}(1-p^{-^1})\alpha_{ij}^{kl},
\ee
\be\label{t22}
\beta_{ij}^{kl}\alpha_{mn}^{kl}=\alpha_{ij}^{kl}\beta_{mn}^{kl}\,,~~~
\forall\, i,j,k,l,m,n;
\ee
and  
\be\label{t23}
\frac{\mu_2(1+|\mu_1|^2)}{z-\mu_1 \mu_2 \bar{z}}\in [0,1],
\ee
where $z=e^{i \theta}\bar{z}$, $z=\mu_2-\mu_1\neq 0$,
$\mu_1$ and $\mu_2$ are the roots of the equation $\alpha^{kl}_{ij}
\lambda^2+\beta^{kl}_{ij}\lambda+\gamma^{kl}_{ij}=0$, for some 
$i,j,k,l$ such that $\alpha^{kl}_{ij} \neq 0$.
This criterion allows one to judge the separability of $\rho$ by
simply calculating its two orthonormal eigenvectors.
Here by using the theorem 1 in \cite{pre},
a simple and effective alternative criterion, 
the negative-partial-trace-criterion, can be also applied in this
case, though
it is not sufficient for general high dimensional case\footnote{
We would like to thank the referees for introducing us the reference}.

In \cite{ent3} we have shown that
if $AA^\dag$ has only two non-zero eigenvalues,
the entanglement of formation of the corresponding pure
state is a monotonically increasing function of the generalized
concurrence. From this the entanglement of formation for a class of
$16\times 16$ mixed density matrices is calculated.

The above approach can be extended to the case of 
multiquantum (particle) systems. We
consider now the entanglement of three $N$ dimensional quantum systems.
A general quantum system is then of the form,
\begin{equation}
\Psi_3=\sum_{i,j,k=1}^N a_{ijk}e_i\otimes e_j\otimes e_k,
\end{equation}
where $a_{ijk}\in\Cb$ and $\sum_{i,j,k=1}^N a_{ijk}a_{ijk}^\ast=1$.
We have one quadratic and three biquadratic invariants:
$$
\ba{ll}
I_0=\displaystyle\sum_{i,j,k=1}^N a_{ijk}a_{ijk}^\ast\,,&
I_1=\displaystyle\sum a_{ijk}a_{ijm}^\ast a_{pqm}a_{pqk}^\ast\,,\\[4mm]
I_2=\displaystyle\sum a_{ikj}a_{imj}^\ast a_{pmq}a_{pkq}^\ast\,,~~~~~~~~~&
I_3=\displaystyle\sum a_{kij}a_{mij}^\ast a_{mpq}a_{kpq}^\ast\,.
\ea
$$
$I_1$ is associated with the exchange of the third sub index. The
corresponding contribution to entanglement is given by
$$
\ba{rcl}
I_0^2-I_1&=&\displaystyle
\sum a_{ijk}a_{pqm}(a_{ijk}^\ast a_{pqm}^\ast-a_{ijm}^\ast
a_{pqk}^\ast)\\[4mm]
&=&\displaystyle\frac{1}{2}
\sum (a_{ijk}a_{pqm}-a_{ijm}a_{pqk})
(a_{ijk}^\ast a_{pqm}^\ast-a_{ijm}^\ast a_{pqk}^\ast)\\[4mm]
&=&\displaystyle\frac{1}{2}
\sum \vert a_{ijk}a_{pqm}-a_{ijm}a_{pqk}\vert^2\,.
\ea
$$
Similarly we have
$$
\ba{rcl}
I_0^2-I_2&=&\displaystyle\frac{1}{2}
\sum \vert a_{ijk}a_{pqm}-a_{iqk}a_{pjm}\vert^2\,,\\[4mm]
I_0^2-I_3&=&\displaystyle\frac{1}{2}
\sum \vert a_{ijk}a_{pqm}-a_{pjk}a_{iqm}\vert^2\,.
\ea
$$
A generalized concurrence can be defined to be
\be\label{cn3}
\ba{rcl}
C_N^3&=&\sqrt{\displaystyle\frac{N}{3(N-1)}(3I_0^2-I_1-I_2-I_3)}\\[3mm]
&=&\displaystyle\sqrt{\frac{N}{6(N-1)}
\sum (\vert a_{ijk}a_{pqm}-a_{ijm}a_{pqk}\vert^2+
\vert a_{ijk}a_{pqm}-a_{iqk}a_{pjm}\vert^2+
\vert a_{ijk}a_{pqm}-a_{pjk}a_{iqm}\vert^2})\,.
\ea
\ee

It is clear that $C_N^3$ is zero when $\Psi_3$ is factorizable, i.e.,
$a_{ijk}=a_i b_j c_k$ for some $a_i, b_j, c_k\in \Cb$. For a maximally
entangled state like $a_{iii}=\frac{1}{\sqrt{N}}$, $i=1,...,N$, and the rest
$a_{ijk}$ being zero, we get $C_N^3=1$.
Nevertheless, when one quantum system
is separated from the other two, e.g., $a_{ijk}=a_{ij}b_k$ for some
$a_{ij},b_k\in \Cb$, $C_N^3$ is not zero, as the three quantum systems
still have some degree of entanglements. In this case,
$$
C_N^3=\sqrt{\frac{N}{6(N-1)}
\sum (\vert a_{ijk}a_{pqm}-a_{iqk}a_{pjm}\vert^2+
\vert a_{ijk}a_{pqm}-a_{pjk}a_{iqm}\vert^2})<1\,.
$$ 
For instance, for a
system of two maximally entangled qubits and one separated qubit,
$a_{111}=a_{222}=a_{112}=a_{221}=\frac{1}{2}$ and the rest $a_{ijk}=0$,
we have $C_N^3=\sqrt{\frac{5}{6}}$.

For $M$ $N$-dimensional quantum systems,
\begin{equation}
\Psi_M=\sum_{i_1,...,i_M=1}^N a_{i_1,...,i_M}e_{i_1}
\otimes ...\otimes e_{i_M},
\end{equation}
$a_{i_1,...,i_M}\in\Cb$,
besides a quadratic invariant:
\be\label{i0m}
I_0=\displaystyle\sum_{i_1,...,i_M=1}^N a_{i_1,...,i_M}
a_{i_1,...,i_M}^\ast\equiv 1,
\ee
there are biquadratic invariants of the form
\be\label{iab}
I_{\alpha\beta}=\displaystyle
\sum a_{\alpha\beta}a_{\alpha\beta^\prime}^\ast 
a_{\alpha^\prime\beta^\prime}a_{\alpha^\prime\beta}^\ast\,,
\ee
where $\alpha$ and $\alpha^\prime$ (resp. $\beta$ and $\beta^\prime$)
are subset of the subindices of $a$, associated to the same sub Hilbert
spaces but with different summing indices.
$\alpha$ (or $\alpha^\prime$) and $\beta$ (or $\beta^\prime$) span the
whole space of a given subindex of $a$.

Under local unitary transformation, $a_{i_1,...,i_M}$ is mapped to
$\sum a_{j_1,...,j_M} b^1_{j_1 i_1}...b^M_{j_M i_M}$, with
$b^k_{j_k i_k}$,
$k=1,...,M$, standing for the unitary transformation on $k$-th quantum
space, $\displaystyle\sum_{l=1}^N 
b^k_{j l}b^{k\ast}_{j^\prime l}=\delta_{jj^\prime}$.
It is straightforward to check that under this transformation
$I_{\alpha\beta}$ is an invariant. 

From (\ref{i0m}) and (\ref{iab}) we have
$$
I_0^2-I_{\alpha\beta}
=\displaystyle\frac{1}{2}
\sum_{\{\alpha,\alpha^\prime,\beta,\beta^\prime\}}^N
\vert a_{\alpha\beta}a_{\alpha^\prime\beta^\prime}- 
a_{\alpha\beta^\prime}a_{\alpha^\prime\beta}\vert^2\,.
$$
Altogether we have $d=2^{M-1}-1$ biquadratic invariants, corresponding to
different selections of the sub index sets of $\alpha$, $\beta$.
The generalized concurrence is then given by
\be\label{cnm}
C_N^M=\sqrt{\frac{N}{d(N-1)}(d I_0^2 - I_1-...-I_d)}
=\sqrt{\displaystyle\frac{N}{2d(N-1)}\sum_p
\sum_{\{\alpha,\alpha^\prime,\beta,\beta^\prime\}}^N
\vert a_{\alpha\beta}a_{\alpha^\prime\beta^\prime}- 
a_{\alpha\beta^\prime}a_{\alpha^\prime\beta}\vert^2}\,,
\ee
where $\displaystyle\sum_p$ stands for the summation over all possible
combinations of the indices of $\alpha$ and $\beta$. From (\ref{cnm})
the separability conditions for multipartite mixed states can be
studied \cite{qsepm}.

We have studied the quantum entanglements
for $N$-dimensional bipartite quantum systems and
multiparticle systems in terms of
of invariants under local unitary transformations.
Using the properties of the
generalized concurrence the entanglement of formation
and the separability of high dimensional mixed states
can be investigated.

\vspace{2.5ex}

{\raggedleft
ACKNOWLEDGEMENTS: We would like to thank C. Bennet, 
L.M. Duan, L.D. Gottesman,
H.K. Lo, R.F. Werner and W.K. Wootters for
useful discussions and communications.}

\end{document}